\documentclass[12pt]{article}
\usepackage{a4wide,graphics,graphicx,amsmath,amssymb,cite,nicefrac}
\usepackage[verbose]{wrapfig}
\usepackage{authblk,hyperref}
\usepackage{url,amsfonts}
\catcode`@=11 \@addtoreset{equation}{section} \catcode`@=12

\newcommand*\DAlambert{\mathop{}\!\mathbin\Box}

\begin{document}

\begin{titlepage}%1
\begin{center}
%\hfill DFPD/2017/TH/\\

\vskip 1.0cm

{\bf \huge The non-Abelian T-dual of \\ \vskip 5pt  Klebanov-Witten Background \\ \vskip 5pt and its Penrose Limits}

\vskip 2.0cm

{\bf \large Sourav Roychowdhury${}^{1}$ and Prasanta K. Tripathy${}^{2}$}

\vskip 30pt
 {\it ${}^1$Chennai Mathematical Institute, \\
SIPCOT IT Park, Siruseri 603 103, India}\\\vskip 5pt
{\it ${}^2$%
Department of Physics, Indian Institute of
Technology
Madras,  \\ Chennai
600 036, India}\\

\vskip 10pt

\texttt{%
souravroy@cmi.ac.in},
\texttt{%
prasanta@iitm.ac.in}

\end{%
center}

\vskip 1.5cm

\begin{%
center} {\bf ABSTRACT}\\[3ex]\end{center}%

In this paper we consider both Abelian as well as non-Abelian T-duals of the Klebanov-Witten background and inspect their various Penrose limits. 
We show that these backgrounds admit pp-wave solutions in the neighbourhood of appropriate null geodesics. We study the quantization of closed 
string propagating on some of the resulting pp-wave backgrounds and comment on the probable field theory duals.

%\today

%\end{center}%

%\noindent

%
\vfill

%\July 2008

\end{titlepage}

\newpage % \setcounter{page}{1} \numberwithin{equation}{section}
%\pagestyle{plain}
%\tableofcontents

\section{\label{Intro}Introduction}

Duality plays a significant role in understanding various aspects of string theory. T-duality is one such example which relates low energy effective
actions of various string theories among each other \cite{Buscher:1987sk,Buscher:1987qj}. One of the most familiar description of T-duality widely 
used in the literature concerns with $U(1)$ isometry. In this case the duality is not merely relating the low energy theories among each other but 
manifests as a symmetry of the full string theory \cite{Rocek:1991ps}. A non-trivial generalization of this duality exists for isometries admitting the 
structure of a non-Abelian group \cite{delaOssa:1992vci}. However, unlike their Abelian counterparts, these non-Abelian T-dualities are not extended 
to full string theory \cite{Giveon:1993ai}. Instead, they are used to relate the low energy theories among each other.

Several aspects of the non-Abelian T-duality have been investigated in recent years. An important development in this area was to generalize
this formalism in the presence of RR fields  \cite{Sfetsos:2010uq}. This in turn was  used as a solution generating technique in  supergravity to obtain new backgrounds.
To demonstrate the applicability of this formalism, a  $SU(2)$ sub-group of  isometry in the near horizon limit of coincident $D3$ branes as well as $D1-D5$ system has been used
to generate a new background in massive type $IIA$ supergravity \cite{Sfetsos:2010uq}. An immediate generalization of this construction for 
non-Abelian T-duals in coset geometries has been carried out \cite{Lozano:2011kb}. Non-Abelian T-duality for the Plich-Warner background
has been carried out in \cite{Dimov:2015rie} where the transformation rules for the  background RR fields were derived from Fourier-Mukai transform. 
These techniques have been applied over and again to 
generate several new backgrounds, as well as to relate known backgrounds among each other \cite{Whiting}. Moreover, the role of non-Abelian 
T-duality in the context of AdS/CFT correspondence has been explored \cite{Itsios:2012zv,Barranco:2013fza,Kooner:2014cqa,Itsios:2013wd,Araujo:2015npa,Macpherson:2014eza}. 
Interesting connections of these dual geometries with the Penrose limit \cite{PenroseLimit} of some well known supergravity backgrounds has been 
established in this context \cite{Lozano:2016kum,Lozano:2016wrs,Dimov:2016rff}.

For a large class of supergravity backgrounds the Penrose limit gives rise to pp-wave geometry. There has been immense study of the
field theory duals for various pp-wave backgrounds during the last two decades \cite{Sadri:2003pr}. It has been proven that the pp-wave solutions provide
exact backgrounds to all orders in $\alpha'$ and $g_s$ in string theory \cite{Amati:1988sa,Horowitz:1989bv}. Thus they have become instrumental in the context of AdS/CFT
correspondence to construct interacting string states from the perturbative gauge theory \cite{Berenstein:2002jq}. More recently, the Penrose limits of non-Abelian
T-dual for the orbifolds of $AdS_5\times S^5$ have been studied in detail \cite{Itsios:2017nou}. Plane wave geometry has been obtained by considering the
Penrose limit along appropriate null geodesic and quantization of string theory in the background of this plane wave geometry has been
carried out and the corresponding field theory dual has been constructed. 

Blowing up the singularities of  orbifolds of $S^5$ gives rise to smooth geometries. One such geometry which has played an interesting
role in understanding the AdS/CFT duality is $T^{1,1}$. This geometry arises as the near horizon limit of coincident $D3$ branes placed
on a conifold singularity \cite{Klebanov:1998hh}. The field theory dual for $AdS_5\times T^{1,1}$ background was first constructed
by Klebanov and Witten to obtain ${\cal N}=1$ SYM theory  \cite{Klebanov:1999tb}. Penrose limit for this background and its field theory
dual were also analysed in detail \cite{Itzhaki:2002kh,Gomis:2002km,PandoZayas:2002dso}. In the present work, we extend the aforementioned 
results about the non-Abelian T-duality, for the Klebanov-Witten background. We consider both Abelian as well as non-Abelian T-dual geometries 
and analyze Penrose limits for various null geodesics. We show that these backgrounds give rise to pp-wave geometries for suitably chosen geodesics. 
We discuss quantization of closed strings propagating in some of these pp-wave backgrounds and comment on the resulting field theory duals. In the 
following we will first summarise the important results discussing Penrose limits of the dual backgrounds obtained from $AdS_5\times S^5$. We 
subsequently consider the generalisation of these results to the Klebanov-Witten background. Finally, we comment on the probable field theory duals  for 
some of the resulting pp-wave backgrounds.

\section{Dual Backgrounds From $AdS_5\times S^5$}

We will first consider the Penrose limits from T-duals of $AdS_5\times S^5$ background. The background metric is given as
\begin{eqnarray}
ds^2 &= & 4 L^2 \big( - \cosh^2 r dt^2 + dr^2 + \sinh^2r d\Omega_3^2 + d\alpha^2 + \sin^2\alpha d\beta^2\big) \cr
&+& L^2 \cos^2\alpha \big(d\theta^2 + d\phi^2 + d\psi^2 + 2 \cos\theta d\phi d\psi\big) \ .
\end{eqnarray}
Here $L$ is the AdS radius and $d\Omega_3$ is the round metric on $S^3$. This background is supported by a self-dual five form field
strength $F_5$. The Penrose limit of this background has been considered in the seminal paper \cite{Berenstein:2002jq}. The field theory dual of the resulting
pp-wave background corresponds to the BMN sector of ${\cal N}=4$  Super-Yang-Mills theory.

The Abelian T-duality along the $\psi$ direction \cite{Lozano:2016kum}, after appropriate coordinate redefinitions, gives rise to the metric 
\begin{eqnarray}
ds^2 = 4 L^2 ds^2({\rm AdS}_5) + 4 L^2\ d\Omega^2_2(\alpha,\beta) + \frac{L^2 d\psi^2}{\cos^2\alpha} + L^2 \cos^2\alpha\ d\Omega^2_2(\chi,\xi) \ ,
\end{eqnarray}
along with a dilaton $\phi$, $B_2$ and a three form field $C_3$.
Here $ds^2({\rm AdS}_5)$ is the metric on ${\rm AdS}_5$ and $d\Omega^2_2(\theta,\phi) = d\theta^2 + \sin^2\theta d\phi^2$ is the metric on $S^2$.
In this background, for motion along the $\xi$ direction the null geodesics are $\{\alpha=0,\chi = \pi/2\}$ and $\{\alpha=\pi, \chi = \pi/2\}$. Focusing in the
vicinity of both these geodesics one gets a pp-wave geometry \cite{Itsios:2017nou}. In addition, pp-waves are also obtained by considering the geodesic $\{\alpha =0, \chi = \pi/2\}$
for motion along $\psi$ and $\xi$ directions. The authors of \cite{Itsios:2017nou} considered quantization of closed string propagating in this background. However their main
focus of discussion was the pp-wave solutions originating from the non-Abelian T-duals.

After T-dualizing along an $SU(2)$ direction \cite{Lozano:2016kum}, the geometry becomes
\begin{eqnarray}
ds^2 = 4 L^2 ds^2({\rm AdS}_5) + 4 L^2 d\Omega^2_2(\alpha,\beta) + \frac{\alpha'^2d\tilde\rho^2}{L^2\cos^2\alpha}
+ \frac{\alpha'^2 L^2 \tilde\rho^2\cos^2\alpha}{\alpha'^2\tilde\rho^2 + L^4\cos^4\alpha} d\Omega^2_2(\chi,\xi) \ .
\end{eqnarray}
In this case, however, the motion along the $\xi$ direction does not admit any pp-wave solution in the vicinity of any of the null geodesics. To obtain
pp-wave solution we need to focus on motion along $\rho (\equiv \alpha' \tilde\rho/L^2)$ and $\xi$ direction \cite{Itsios:2017nou}. In this case, the null 
geodesic is located at $\{\chi = \pi/2, \alpha = 0\}$.
The Lagrangian for a massless particle moving on a null geodesic admits two cyclic coordinates giving rise to the conservation of energy and
angular momentum. Expanding around the null geodesic and making appropriate coordinate redefinition, one can bring the resulting pp-wave
metric to the standard Brinkmann form \cite{Itsios:2017nou}:
\begin{equation}
\label{ds2ppNATD}
 \begin{aligned}
   ds^2 & = 2 \, du \, dv + d\bar{r}^2 + \bar{r}^2 \, d\Omega^2_3 + dx^2 + x^2 \, d\beta^2 +  dz^2 + dw^2
   \\[5pt]
   & - \Bigg[   \frac{\bar{r}^2}{16} + \frac{x^2}{16} \,  (8 J^2 - 1) + \frac{ (\rho^2 + 1)^2}{\rho^4} J^2 z^2 - F_z \, z^2 - F_w \, w^2  \Bigg] \, du^2 \ ,
 \end{aligned}
\end{equation}
where
\begin{equation}
  F_z = \frac{4 \, J^2 \big(   4 \, \rho^2 + 1  \big) + 3 \, \big(   4 \, J^2 - 1  \big) \, \rho^4}{4 \, \rho^4 \, \big(   \rho^2 + 1  \big)^2} \ ,
\qquad  F_w = - \frac{3}{4 \, \big(  \rho^2 + 1  \big)^2} \ .
\end{equation}
The NS-NS three form $H_3$ and RR four form $F_4$ hold the following expressions upon taking the Penrose limit:
\begin{equation}
 H_3= \frac{1}{2} \, \frac{\rho^2 + 3}{\rho^2 + 1} \, du \wedge dz \wedge dw \; ,
 F_4  = \frac{2 \, J \, x \, \sqrt{\rho^2 + 1}}{\tilde{\tilde g}_s} \, du \wedge dx \wedge dz \wedge d\beta \ .
\end{equation}
The authors of  \cite{Itsios:2017nou} studied propagation of closed strings in this background. Solutions to the equation of motion are constructed. Further, they
have proposed a field theory dual for this background.

\section{Abelian T-dual of Klebanov-Witten Background}

Our goal in the present work is to generalize these results for the  background $AdS_5\times T^{1,1}$. This geometry corresponds to the near horizon limit
of parallel $D3$ branes at  conical singularities, and provides one of the earliest examples of the AdS/CFT correspondence. The metric corresponding to the
geometry has the form
\begin{equation}
ds^2=L^2 ds^2_{AdS_{5}}+L^2 ds^2_{T^{1,1}},
\end{equation}

\begin{equation}
ds^2_{AdS_{5}}=-\cosh^2r\ dt^2 + dr^2 + \sinh^2r\ d\Omega^2_{3},
\end{equation}

\begin{equation}
ds^2_{T^{1,1}}= \lambda_{1}^2\ d\Omega^2_{2} \big(\theta_{1},\phi_{1}\big) + \lambda_{2}^2\ d\Omega^2_{2} \big(\theta_{2},\phi_{2}\big)
+ \lambda^2\big(d\psi + \cos\theta_{1} d\phi_{1} + \cos\theta_{2} d\phi_{2}\big)^2 \ .
\end{equation}
Here $L$ is the $AdS_5$ radius, and the parameters $\lambda,\lambda_1,\lambda_2$ in the $T^{1,1}$ metric have the numerical  values
  $\lambda_{1}^2=\lambda_{2}^2=\frac{1}{6}, \lambda^2=\frac{1}{9}$. In addition, the supergravity background contains a constant dilation $\Phi$,
and a self-dual RR five form field strength
\begin{equation}
F_{5}= \frac{4}{g_{s}L}\big[{\rm Vol} (AdS_{5}) - L^5 {\rm Vol}(T^{1,1})\big] .
\end{equation}

We will study both Abelian as well as non-Abelian T-duals of this background. We will first focus on the Abelian T-duality. The brane constructions for the
corresponding dual geometry was studied in detail \cite{Uranga:1998vf,Dasgupta:1998su}. They correspond to various intersecting branes in type IIA string theory. In the following we will consider
the Penrose limits for the Abelian T-duality of this background about some of the $U(1)$ isometry directions. %$\psi$ as well as $\phi_2$-directions.

The background has a manifest $U(1)$ invariance along $\phi_1,\phi_2$ and $\psi$ directions. First focus on the azimuthal directions $(\phi_1,\phi_2)$. There is a symmetry 
under the exchange of $(\theta_1,\phi_1)$  with $(\theta_2,\phi_2)$. Thus, it would be sufficient to consider the duality along one of these two directions. Here we will 
consider the Abelian T-duality along $\phi_2$ isometry. It is straightforward to obtain the dual geometry using the standard rules of T-duality \cite{Bergshoeff:1995as}. 
The duality preserves all the supersymmetries of the Klebanov-Witten  background.The 
field theory dual corresponding to this background has been analysed \cite{Itsios:2017cew}. The metric corresponding to the dual background is given by
\begin{equation}\label{tdab}
L^{-2} d\hat{s}^2=  ds^2_{AdS_{5}} +  \lambda_{1}^2  \left[d\Omega_{2}^2\big(\theta_{1},\phi_{1}\big) + d\theta_{2}^2
+ \frac{\lambda^2 \sin^2 \theta_2}{P(\theta_2)} \big(d\psi + \cos \theta_{1} d\phi_{1}\big)^2 + \frac{d\phi_{2}^2}{\lambda_{1}^2 P(\theta_{2})}\right] .
\end{equation}
Here we have used the notation $P(\theta_{2})= \lambda^2 \cos^2\theta_2 +  \lambda_{2}^2 \sin^2\theta_2 \ .$
The dilaton and the NS-NS two form fields are given respectively by
\begin{equation}
e^{-2 \hat{\Phi}}= \frac{L^2}{g_{s}^2} P(\theta_{2})  \ ,
\end{equation}
and
\begin{equation}
\hat{B}_{2}= - \frac{L^2 \lambda^2 \cos\theta_2}{P(\theta_{2})} \ \Big(d\phi_{2} \wedge d\psi + \cos\theta_1 d\phi_{2} \wedge d\phi_{1}\Big) \ ,
\end{equation}
The RR two form $F_2$ vanishes, whereas the RR four form $F_4$ has the expression
\begin{equation}
\hat{F}_{4}= \frac{4L^4 \lambda \lambda_{1}^4}{g_{s}} \sin\theta_1 \sin\theta_2 \ d\theta_1 \wedge d\phi_1 \wedge d\theta_2 \wedge d\psi \ .
\end{equation}

We will now focus on obtaining Penrose limits for this background. To this end, consider the geodesic equation
\begin{equation}\label{geod}
\frac{d^2 x^{\mu}}{du^2}+\Gamma^{\mu}_{\nu\rho} \frac{dx^{\nu}}{du} \frac{dx^{\rho}}{du} =0 \  .
\end{equation}
Here $\{x^\mu\}$ are the space-time coordinates and $u$ denotes the affine parameter along the geodesic. We will consider
the motion along some isometry direction. If $x^{\mu_0}$ is such an isometry direction, then the velocity as well as acceleration along
any $x^\mu, \mu\neq\mu_0$ vanish:
\begin{equation}
\frac{d x^\mu}{du} = 0 = \frac{d^2 x^\mu}{du^2}  , \mu\neq \mu_0 \ .
\end{equation}
Substituting the above in \eqref{geod}, we find
\begin{equation} \label{abeliangeod}
\partial^\mu g_{\mu_0\mu_0} = 0 \ .
\end{equation}
To obtain the Penrose limit, we need to focus in the vicinity of null geodesics. Thus, in addition to the above condition, we must
require $ds^2=0$.

We will now analyse the motion along various isometry directions of the T-dual geometry \eqref{tdab} and obtain the corresponding Penrose limits. 
Consider first the $\phi_1$ isometry. The geodesic equation along this direction is $\partial_\mu g_{\phi_1\phi_1} = 0$. The relevant component of the metric is
\begin{equation}
g_{\phi_{1}\phi_{1}}= L^2\ \Big[\lambda_{1}^2 \sin^2\theta_1 + \frac{\lambda_{1}^2 \lambda^2\sin^2\theta_2}{\lambda^2 \cos^2\theta_2+ \lambda_{2}^2 \sin^2\theta_2} \cos^2\theta_1\Big] \ .
\end{equation}
The geodesic condition  for $\mu = \theta_1$ as well as for $\mu=\theta_2$  can be solved to obtain $\theta_{1}=(0, \frac{\pi}{2}, \pi)$ and  $ \theta_{2}=(0, \frac{\pi}{2}, \pi)$
respectively. This gives us four geodesics: $\{\theta_1=0,\theta_2=\pi/2\}, \ \{\theta_1=\pi ,\theta_2= \pi/2 \}, \ \{\theta_1= \pi/2 , \theta_2=0\}$ and 
$\{\theta_1= \pi/2, \theta_2=\pi \}$. We first consider the following large $L$ expansion  around the geodesic $\{\theta_{1}=0, \theta_{2}=\frac{\pi}{2}\}$: 
% with following expansions
%
\begin{equation}
r=\frac{\bar{r}}{L}, \ \theta_{1}=\frac{z}{L},  \ \theta_{2}=\frac{\pi}{2}+\frac{x}{L}, \ t=ax^{+}, \ \phi_{1}=bx^{+} + \frac{x^{-}}{L^2},\ \phi_{2}=\frac{\phi_{2}}{L},
\end{equation}
while keeping $\psi$ unchanged. Here $a$ and $b$ are unknown parameters.
Ignoring the subleading terms in $\L\rightarrow\infty$ limit,  we find the T-dual metric to have the following expression
\begin{eqnarray}
ds^2 &=& d\bar{r}^2 + \bar{r}^2 d\Omega_{3}^2 + \lambda_{1}^2 dz^2 + \lambda _{1}^2 dx^2 - \Big[\bar{r}^2 a^2+ b^2 z^2 \big(\lambda^2-\lambda_{1}^2\big)\Big](dx^{+})^2\nonumber\\
 &-& \lambda^2 \Big[bz^2 d\psi dx^+ -2d\psi dx^- - 2b dx^+ dx^-\Big] -\frac{\lambda^4}{\lambda_{2}^2} x^2 \big(d\psi + b dx^+ \big)^2 + \frac{1}{\lambda_2^2} d\phi_{2}^2\nonumber\\
 &-& L^2 \Big[a^2 (dx^{+})^2 - \lambda^2 \big(d\psi + b dx^+ \big)^2 \Big] \ .
\end{eqnarray}
Note that the metric diverges in the limit $L\rightarrow\infty$ due to the presence of $O(L^2)$ terms. This divergence occurs because, in this case we have 
not been able to impose the geodesic to be null. This amounts to setting 
$$ a^2 (dx^{+})^2 - \lambda^2 \big(d\psi + b dx^+ \big)^2 = 0 \ . $$
Clearly, this can't be satisfied for any choice of the parameters $a$ and $b$ due to the presence of the $d\psi$ term. A similar analysis
can be carried out for the geodesic $\{\theta_1=\pi ,\theta_2= \pi/2 \}$, leading to a divergent metric in the large $L$ expansion.

In contrast, expanding the T-dual metric around the remaining two geodesics gives rise to pp-wave geometry as we will show currently. Consider first the following expansion
about the geodesic $\{\theta_1= \pi/2 , \theta_2=0\}$: 
\begin{equation}
r=\frac{\bar{r}}{L} \ , \ \theta_{1}=\frac{\pi}{2}+\frac{z}{L} \ ,  \ \theta_{2}=\frac{x}{L} \ , \ t=ax^{+} , \ \phi_{1}=bx^{+} + \frac{x^{-}}{L^2} \ ,\ \phi_{2}=\frac{\phi_{2}}{L} \ ,
\end{equation}
keeping the $\psi$-coordinate unchanged. Here, as before $a$ and $b$ are unknown parameters to be chosen suitable\footnote{To get the metric
in standard from, we set $a=1/\lambda_1, b = 1/\lambda_1^2$ and, in addition, we rescale some of the coordinates as $x\rightarrow \sqrt{6}x,
 \ z\rightarrow \sqrt{6}z, \ \phi_2 \rightarrow \frac{1}{3} \phi_2$.}
in order to obtain
\begin{equation}\label{abppwave}
ds^2_{pp}= 2dx^+ dx^- + d\bar{r}^2 + \bar{r}^2 d\Omega_{3}^{2} + dz^2 + dx^2 + x^2 d\psi^2 + d\phi_{2}^2 - 6 \big(\bar{r}^2+ 6 z^2-6x^2 \big)(dx^+)^2 \ .
\end{equation}
Clearly, the background geometry is  a pp-wave solution in the standard Brinkmann form.
The background dilaton has the expression
\begin{equation}
e^{-2\hat{\Phi}}=\frac{1}{\tilde{g}_{s}^2} \lambda^2 \ ,
\end{equation}
and NS-NS two-form field
\begin{equation}
\hat{B}_{2}= 2\sqrt{6} z \ d\phi_{2} \wedge dx^+\ ,
\end{equation}
with corresponding field strength
\begin{equation}
\hat{H}_{3}= 2\sqrt{6} \ dz\wedge d\phi_2 \wedge dx^+ \ .
\end{equation}
The RR fields in this limit has the expression
\begin{equation}
\hat{F}_{2}=0 \ , \ \hat{F}_{4}= \frac{4\sqrt{6}}{3\tilde{g}_{s}}\ x\ dz \wedge dx^+ \wedge dx \wedge d\psi \ .
\end{equation}
Taking Penrose limit for the geodesic $\{\theta_1= \pi/2, \theta_2=\pi \}$  also leads to a pp wave geometry with the same metric as \eqref{abppwave}. The expressions
for the background fields are also quite similar. We omit the details because the analysis is identical to the aforementioned discussion.

Now consider motion along the $\phi_2$ direction. To obtain the geodesics along this isometry, consider the metric component
\begin{equation}
g_{\phi_{2}\phi_{2}}= \frac{L^2}{\lambda^2 \cos^2\theta_2+ \lambda_{2}^2 \sin^2\theta_2} \ .
\end{equation}
From the geodesic condition, $\partial_{\theta_{2}} g_{\phi_{2}\phi_{2}}=0$ we find $ \theta_{2}=\big(0, \pi/2, \pi \big)$.
Consider the following expansion around the geodesic $\{\theta_1=0,\theta_2=0\}$:
\begin{equation}
r=\frac{\bar{r}}{L}, \ \theta_{1}=\frac{z}{L},  \ \theta_{2}=\frac{x}{L}, \ t=ax^{+}, \ \phi_{2}=bx^{+} + \frac{x^{-}}{L^2},
\end{equation}
keeping $\phi_1$ and $\psi$ unchanged. To remove the $O(L^2)$ divergent piece in the metric, we need to choose $a=\lambda, b=\lambda^2$ . 
This choice leads to a null geodesic. With appropriate redefinition of the $x$ and $z$ coordinates, we find the T-dual metric as
\begin{equation}
ds^2= 2dx^+dx^- +d\bar{r}^2 + \bar{r}^2 d\Omega_{3}^{2} + dz^2 +z^2d\phi_{1}^2 + dx^2 + x^2 (d\psi+d\phi_1)^2 - \frac{1}{9} \big(\bar{r}^2 + 3x^2\big) (dx^+)^2.
\end{equation}
Though the metric is now finite, the scalar curvature for this solution is non-zero and hence it does not correspond to a pp-wave geometry.
This is due to the fact that the geodesic is placed on a singular location. The metric component $g_{\phi_1\phi_1}$ vanishes for the values $\{\theta_1=0,\theta_2=0\}$. 
This is a generic feature and hence we will no longer consider such singular geodesics from now on.

Finally, consider the $\psi$-isometry direction. The null geodesics can be obtained by considering the $g_{\psi \psi}$ component
of the metric, which  is given by
\begin{equation}
g_{\psi \psi}=L^2 \frac{\lambda_{1}^2 \lambda^2\sin^2\theta_2}{\lambda^2 \cos^2\theta_2+ \lambda_{2}^2 \sin^2\theta_2} \ .
\end{equation}
Solving the geodesic condition one obtains $ \theta_{2}=\big(0, \pi/2, \pi \big)$. For  the values $\theta_{2}=(0, \pi)$ the above metric component vanishes
and hence, we do not consider these values here. Consider the following expansion around the geodesic $\theta_{1}=0$ and  $\theta_{2}=\frac{\pi}{2}$:
\begin{equation}
r=\frac{\bar{r}}{L}, \ \theta_{1}=\frac{x}{L},  \ \theta_{2}=\frac{\pi}{2}+\frac{z}{L}, \ t=ax^{+}, \ \psi =bx^{+} + \frac{x^{-}}{L^2},\ \phi_{2}=\frac{\phi_{2}}{L},
\end{equation}
while keeping the $\phi_1$ coordinate unchanged. The leading terms of T-dual metric in the limit $L\rightarrow\infty$ are given by
\begin{eqnarray}
ds^2 &=& d\bar{r}^2 + \bar{r}^2 d\Omega_{3}^2 + \lambda_{1}^2 dx^2 + \lambda _{1}^2 dz^2 + \Big[\Big(\lambda_{1}^2-\lambda^2\Big)x^2-\frac{\lambda^4}{\lambda_{2}^2}z^2\Big] d\phi_{1}^2 + \frac{1}{\lambda_{2}^2}d\phi_{2}^2 \cr
&-&\Big[\bar{r}^2 a^2+ \frac{\lambda^4}{\lambda_{2}^2}b^2 z^2 \Big](dx^{+})^2%\nonumber\\
+ \lambda^2 \Big[2b dx^+ dx^- -b\Big(x^2+ \frac{\lambda^2}{\lambda_{2}^2}2z^2\Big)dx^+ d\phi_1 +2 dx^- d\phi_1\Big] \nonumber\\
 &-& L^2 \Big[a^2 (dx^{+})^2 - \lambda^2 \big(b dx^+ + d\phi_1\big)^2 \Big] \ .
\end{eqnarray}
This contains a divergent term which can�t be removed by any choice of the parameters $a$ and $b$. This is because, in this case too, we do not have a null geodesic
for any choice of the parameters $a$ and $b$. Hence motion along the isometry direction $\psi$ does not lead to any 
pp-wave geometry.

In the above, we have considered the Abelian T-duality along $\phi_2$ direction and analysed all the geodesics admitted by this geometry.  Some of these geodesics 
are singular and taking Penrose limit does not lead to any interesting solution in such cases. Only two of these geodesics 
are null. Taking Penrose limit in the vicinity of these two null geodesics leads to pp wave geometries.  An identical result will hold for 
T-duality along $\phi_1$ direction. We will now focus on the remaining isometry direction $\psi$. Using the standard rules of T-duality\cite{Bergshoeff:1995as}, we obtain
\begin{equation} \label{abelianTdual}
d\hat{s}^2= L^2 ds^2_{AdS_{5}} + L^2\ \Big[\lambda_{1}^2 d\Omega_{2}^2\big(\theta_{1},\phi_{1}\big)
+ \lambda_{2}^2 d\Omega_{2}^2\big(\theta_{2},\phi_{2}\big) + \frac{1}{\lambda^2} d\psi^2\Big] \ .
\end{equation}
Here we have rescaled  $\psi \rightarrow\frac{L^2}{{\alpha}^{\prime}} \psi$ in order to get $L^2$ as a common factor in the metric and set $\alpha'=1$ for
convenience. The resulting T-dual geometry has the well known product form $AdS_5\times S^2\times S^2\times S^1$. Unlike the previous case, the background 
is non-supersymmetric in this case. The NS-NS sector also contains a
constant dilaton
\begin{equation}
e^{-2\hat{\Phi}}= \frac{\lambda^2 L^2}{g_{s}^2} \ ,
\end{equation}
and a two-form field
\begin{equation}
\hat{B}_{2}= - L^2 \ \big[\cos\theta_{1} d\phi_{1}+ \cos\theta_{2} d\phi_{2}\big] \wedge d\psi \ ,
\end{equation}
with field strength
\begin{equation}
\hat{H}_{3}= L^2 \ \big[\sin\theta_{1} d\theta_{1}\wedge d\phi_{1}+ \sin\theta_{2} d\theta_{2}\wedge d\phi_{2}\big] \wedge d\psi \ .
\end{equation}
The RR sector of the resulting background consists of a non-vanishing four-form flux
\begin{equation}
\hat{F}_{4}= \frac{4L^4 \lambda\lambda_{1}^2\lambda_{2}^2}{g_{s}} \sin\theta_{1}\sin\theta_{2} \ d\phi_{1} \wedge d\theta_{1} \wedge d\phi_{2} \wedge d\theta_{2} \  .
\end{equation}

We will now focus on the metric \eqref{abelianTdual}. Clearly $\phi_1,\phi_2$ and $\psi$ are the isometry directions.
Motion along the $\psi$ direction does not give any non-trivial constraint. Since the analysis is identical for both $\phi_1$ and
$\phi_2$, it will be sufficient to consider geodesics along one of these directions. For the $\phi_1$ isometry direction the condition
\eqref{abeliangeod} gives $ \theta_{1}=(0, \frac{\pi}{2}, \pi)$. However, the singular values $\theta_1=0$ and $\pi$ correspond to points
and not curves and hence we do not have any corresponding geodesics. This leaves behind the choice $\theta_1=\pi/2$. To get the
Penrose limit, we consider the following large $L$ expansion of the dual metric \eqref{abelianTdual} retaining the leading terms:
\begin{equation}
r=\frac{\bar{r}}{L}, \ \theta_{1}=\frac{\pi}{2} + \frac{z}{L}, \ \theta_{2}=\frac{x}{L}, \ t=ax^{+}, \ \phi_{1}=bx^{+} + \frac{x^{-}}{L^2},\ \psi=\frac{y}{L}, \ \phi_{2}=\beta ,
\end{equation}
and redefine the string coupling as $g_{s}=L \ \tilde{g_{s}}$ to ensure that the dilaton remains finite. To obtain a null geodesic we must impose the condition $a=\lambda_1 b$.
In addition, we set $\lambda_{1}^2b=1$ and make the co-ordinate redifinitions $x^{+}=u,\ x^{-}=v, \ z\rightarrow\sqrt{6}z, \ x\rightarrow\sqrt{6}x ,
\ y\rightarrow \frac{1}{3}y $ to bring the resulting pp-wave metric to the standard form:
\begin{equation}\label{bgppwave}
ds^2= 2dudv + d\bar{r}^2 + \bar{r}^2 d\Omega_{3}^{2} + dz^2 + dx^2 + x^2 d\beta^2 + dy^2 - 6(\bar{r}^2+ 6 z^2)du^2.
\end{equation}
%
%Interestingly, this geometry coincides with the pp-wave solution obtained from the T-dual of $AdS_5\times S^5$ background.
%
The expressions for the dilaton and NS-NS three-form flux in this limit are given as:
\begin{equation}
e^{-2\hat{\Phi}}=\frac{\lambda^2}{\tilde{g}_{s}^2} \  ,
\end{equation}
and
\begin{equation}
\hat{H}_{3}= 2\sqrt{6} \ dz\wedge du \wedge dy \  .
\end{equation}
Field strengths for the RR fluxes have the expression:
\begin{equation}
\hat{F}_{2}=0, \ \hat{F}_{4}=\frac{4 \sqrt{6}}{3\tilde{g}_{s}} x \ du\wedge dz \wedge d\beta \wedge dx \ .
\end{equation}

%The geodesics we considered so far are all static. 
The null geodesics can also carry angular momentum. To obtain such a geodesic, we consider motion along $\phi_{1}$
and  $\psi$ directions. The geodesic equation now implies that  $\theta_{1}=\pi/2, \theta_{2}= 0$. Consider the Lagrangian
for a massless particle moving along this geodesic:
\begin{equation}
\mathcal{L}= \frac{1}{2} g_{\mu\nu} \dot{X}^{\mu}\dot{X}^{\nu} \ .
\end{equation}
Here we choose $u$ to be the affine parameter and the dots denote derivative with respect to it.
Substituting the explicit expression for the background metric \eqref{abelianTdual}  in the above Lagrangian we find
\begin{equation}
\mathcal{L}= \frac{L^2}{2} \Big(-\dot{t}^2 + \frac{1}{6} \dot{\phi}_{1}^2 + 9 \dot{\psi}^2\Big) \ .
\end{equation}
Clearly, the conjugate momenta corresponding to the generalized coordinates $t,\phi_1$ and $\psi$ are conserved.
Suitably choosing the affine parameter $u$ we set
$$\frac{\partial \mathcal{L}}{\partial \dot{t}}=-L^2 \dot{t}=-L^2 \ . $$
Denoting $J$ to be the conserved quantity associated with the variable $\phi_1$, we have
$$  \frac{\partial \mathcal{L}}{\partial \dot{\phi}_{1}}= \frac{1}{6} \dot{\phi}_{1}L^2=-JL^2  \ . $$
The conserved momentum with respect to the variable $\psi$ however can no longer be arbitrary. It has to be determined
by requiring the geodesic to be null, {\it i.e.}, we set ${\mathcal L} = 0$. We find
\begin{equation}
\dot{\psi}^2=\frac{1}{9} \big(1-6J^2\big) \ , \
\end{equation}
which upon integration gives
$$ \psi= \frac{1}{3} \sqrt{1-6J^2}\ u \ . $$
Here we set the constant of integration to zero. From the above expression we find that in order to get a real value for $\psi$
the angular momentum $J$ must be bounded by
\begin{equation}
0 \leq J \leq \frac{1}{\sqrt{6}} \ .
\end{equation}

To obtain the Penrose limit for a null geodesic carrying angular momentum $J$ on the $(\psi,\phi_1)$ plane
around $r=0, \ \theta_{1}=\frac{\pi}{2}, \ \theta_{2}=0$, we redefine the coordinates
\begin{equation}
r=\frac{\bar{r}}{L}, \ \theta_{1}=\frac{\pi}{2}+ \frac{z}{L}, \ \theta_{2}=\frac{x}{L} \ .
\end{equation}
and consider the following expansion in the limit $\L\rightarrow\infty$:
\begin{equation}
dt= c_{1} du,\ d\phi_{1}= c_{2} du +c_{3} \frac{dw}{L}, \ d\psi= c_{4} du + c_{5} \frac{dw}{L} + c_{6} \frac{dv}{L^2} \ ,
\end{equation}
Requiring the geodesic to be null sets the constant coefficients $c_1,c_2$ and $c_4$ the values
\begin{equation}
c_{1}=1, \ c_{2}=-6J, \ c_{4}= \frac{1}{3} \sqrt{1-6J^2} \ .
\end{equation}
The metric, in addition contains a $\mathcal{O}(L)$ which can be removed upon requiring
 $ \lambda_{1}^2 c_{2}c_{3} + \frac{1}{\lambda^2}c_{4}c_{5}=0$.
 Normalizing the coefficient of $dw^2$  to unity gives the condition $\lambda_{1}^2 c_{3}^2 + \frac{1}{\lambda^2}c_{5}^2=1$.
 Similarly, appropriate  normalization of the cross term $2dudv$ gives $\frac {c_{4}c_{6}}{\lambda^2}=1$.
These condition can be solved uniquely to obtain the remaining coefficients $c_3, c_5$ and $c_6$. We find
\begin{equation}
c_{3}=\sqrt{6(1-6J^2)},\ c_{5}= J \sqrt{\frac{2}{3}}, \ c_{6}= \frac{1}{3} \frac{1}{\sqrt{1-6J^2}}.
\end{equation}

The resulting pp-wave metric after a rescaling $x\rightarrow \sqrt{6} x, z\rightarrow \sqrt{6} z$  has the expression
\begin{equation}\label{ppwave1}
ds^2_{pp}=2 dudv + d\bar{r}^2 + \bar{r}^2 d\Omega_{3}^2 + dz^2 + dx^2 + x^2 d\beta^2 + dw^2 - \big(\bar{r}^2 +36J^2z^2\big)du^2.
\end{equation}
The background dilaton and $B_2$ field are found to be
\begin{equation}\label{pwa1}
\ e^{-2\hat{\Phi}}=\frac{\lambda^2}{\tilde{g}_{s}^2} \ , \ \hat{B}_{2}= 2z\ dw \wedge du + x^2 \sqrt{1-6J^2} \ d\beta \wedge du \ ,
\end{equation}
with the corresponding three form flux
\begin{equation}\label{pwa2}
\hat{H}_{3}= 2\ dz\wedge dw \wedge du + 2x \sqrt{1-6J^2} \ dx \wedge d\beta \wedge du \ .
\end{equation}
In addition, the RR fluxes have the limit
\begin{equation}\label{pwa3}
\hat{F}_{2}=0, \  \hat{F}_{4}= \frac{4\sqrt{6}}{3\tilde{g}_{s}} J x \ du \wedge dz \wedge dx \wedge d\beta \ .
\end{equation}

Before closing this section, we note that all the pp-wave backgrounds we have obtained in this paper do indeed satisfy the supergravity equations.
In the following, we demonstrate this for the background specified by eqs.\eqref{ppwave1}-\eqref{pwa3}. For type-$IIA$ supergravity, the Bianchi 
identity and gauge field equation  are given as 
\begin{eqnarray} \label{binachi}
&& dH_3 = 0 \ , \ dF_{2} = 0 \ , \ dF_{4} = H_{3} \wedge F_{2} \ , \cr
%\end{eqnarray}
%and 
%\begin{eqnarray}
&& d\big(e^{-2\hat\Phi} *H\big) - F_2\wedge *F_4 - \frac{1}{2} F_4\wedge F_4 = 0 \ , \cr
&& d*F_2 + H_3 \wedge *F_4 = 0 \ , \cr
&& d*F_4 + H_3\wedge F_4 = 0 \ .
\end{eqnarray}
A quick inspection of the background shows that the Bianchi identities are indeed satisfied. The equation of motion for $B_2$ is satisfied for our background,
because the dilaton is constant, $F_2=0$ and $F_4\wedge F_4 = 0$. Further, the hodge dual of $H_3$, given by 
$$*H_3 = 2 \big(dx\wedge d\beta +  \sqrt{1-6J^2}\ dz\wedge d\omega\big)\wedge du \wedge d\Omega_4 $$
is closed. To verify the $F_2$ equation of motion, note that $F_2$ is zero and 
$$ *F_4 = \frac{4 \sqrt{6}}{3\tilde g_s} J\ du\wedge d\Omega_4 \ , $$
and hence $H_3\wedge *F_4 = 0$. The last equation holds because $*F_4$ is closed and $H_3\wedge F_4 = 0$.

The equations of motion for metric and dilaton are given as
\begin{eqnarray}
&& R_{\mu\nu} + 2D_{\mu}D_{\nu}\hat\Phi = \frac{1}{4} H_{\mu\nu}^2 + e^{2\Phi} \Bigg[\frac{1}{2} (F_2^2)_{\mu\nu} + \frac{1}{12} (F_4^2)_{\mu\nu} - \frac{1}{4} g_{\mu\nu} \Big( \frac{1}{2} F_{2}^2 + \frac{1}{4!}F_{4}^2 \Big)\Bigg] \ , \nonumber\\
&& R + 4D^2\hat\Phi - 4(\partial\hat\Phi)^2 - \frac{1}{12}H^2=0 \ .
\end{eqnarray}
  To verify these equations, note that 
$\hat \Phi = \text{const}, H^2=F_4^2=0=R$. A straightforward computation shows that only the $uu$-components of $R_{\mu\nu},H^2_{\mu\nu}$ 
and $(F_4^2)_{\mu\nu}$ are non-vanishing. They are given by
\begin{equation}
H^2_{uu} = 16-48 J^2 \ , \ (F_4^2)_{uu} = 64 J^2/{\tilde g_s^2} \ , \ {\rm and} \ R_{uu} = 4 + 36 J^2 \ .
\end{equation}
Substitution the above we can see that the corresponding equations of motion are indeed satisfied.

\section{Quantization of Closed Strings Propagating in the pp-wave Geometry}% in Motion in $\phi_1$ and $\psi$ direction in Abelian T-dual along $\psi$ direction}

In this section we will study the quantization of closed strings propagating in the pp-wave background. We will focus on the pp-wave solution \eqref{ppwave1}
carrying an angular momentum which has been obtained from the dual geometry  by performing an Abelian T-duality along $\psi$-isometry.
The string world sheet action is given by
%
%For a string moves in a background metric $G_{\mu\nu}$ with background fields $B_{\mu\nu}$ and $\Phi$, the action is governed by Polyakov action as
%
\begin{equation}
S=-\frac{1}{4\pi\alpha^{\prime}}\int d\tau d\sigma \Big[\sqrt{g} g^{\alpha\beta}G_{\mu\nu} \partial_{\alpha}X^{\mu}\partial_{\beta}X^{\nu} + \epsilon^{\alpha\beta}B_{\mu\nu} \partial_{\alpha}X^{\mu}\partial_{\beta}X^{\nu} + \alpha^{\prime} \sqrt{g}\ \mathcal{R}^{(2)} \Phi\Big] \ ,
\end{equation}
Here $\{\alpha,\beta\}$ denote the worldsheet coordinates $(\tau,\sigma)$ and $\{\mu,\nu\}$ denote the spacetime coordinates,
$G_{\mu\nu}$ is the background metric, $B_{\mu\nu}$ and $\Phi$ are the background NS-NS two-form and dilaton respectively. We choose
the convention $\epsilon^{\tau\sigma}=-\epsilon^{\sigma\tau}=1$  and gauge fix the worldsheet metric $g_{\alpha\beta}$ such that
$\sqrt{g} g^{\alpha\beta}=\eta^{\alpha\beta}$, with $  -\eta_{\tau\tau}=\eta_{\sigma\sigma}=1 $.
Further, we designate the string coordinates in the manner
\begin{equation}
U=u,\ V=v, \ \big(X^1,X^2,X^3,X^4\big)\in \bar{r},\Omega_{3},\ \big(X^5,X^6\big)\in x,\beta , \ \big(X^7,X^8\big)\in z,w \ ,
\end{equation}
and consider the light cone gauge $U=\tau $ with $p^+=1$ in order to fix the residual diffeomorphism invariance. The worldsheet action
for the pp wave background \eqref{ppwave1} then becomes

\begin{eqnarray}
S=-\frac{1}{4\pi\alpha^{\prime}}\int d\tau d\sigma\ \Big[ \sum_{i=1}^{8}\partial X^{i}.\ \partial X^{i} +  \sum_{i=1}^{4} (X^i)^2
+ X^8 \partial_{\sigma}X^7 - X^7 \partial_{\sigma}X^8 \cr
 -(X^5)^2 \ \sqrt{1-6J^2} \ \partial_{\sigma} X^6
 + 36 J^2 (X^7)^2 \Big] \ ,
\end{eqnarray}
In the above the inner product is defined with $\eta_{\alpha\beta}$ .
The corresponding Euler-Lagrange equations are given as

\begin{equation}\label{uncoupled}
\DAlambert X^{i}-X^{i}=0 \ ,\\ i=1,2,3,4,
\end{equation}

\begin{equation}
\DAlambert X^{5} +   \sqrt{1-6J^2} \ X^5 \ \partial_{\sigma} X^6=0 \ ,
\end{equation}

\begin{equation}
\DAlambert X^{6}  - \sqrt{1-6J^2} \ X^5 \  \partial_\sigma X^5 =0 \ ,
\end{equation}

\begin{equation}
\DAlambert X^{7}-36 J^2 X^7 + \frac{1}{2} \ \partial_{\sigma} X^{8}=0 \ ,
\end{equation}

\begin{equation}
\DAlambert X^{8}-\frac{1}{2} \ \partial_{\sigma} X^{7}=0 \ .
\end{equation}

The first of the above equations, eq.\eqref{uncoupled} is a linear equation  involving the uncoupled fields $X^i, i=1,\ldots,4.$
Considering an ansatz of the form $X \sim e^{-i\omega t + in \sigma}$,
it is straightforward to obtain the frequencies of the respective modes
\begin{equation}
\omega^{2}_{n,i}=n^2 + 1, \\ i=1,\ldots, 4 .
\end{equation}
The last two equations involving $X^7$ and $X^8$ can be decoupled giving rise to two fourth order linear partial differential equations
with corresponding mode frequencies
\begin{equation}\label{eigenfreq}
\omega^{2}_{n,i}=n^2 + \frac{1}{2} \Big[36J^2\pm \sqrt{(36 J^2)^2+n^2} \Big], \\ i=7,8 .
\end{equation}
These modes can be related to the fourth order Pais-Uhlenbeck oscillator as in the case of the Pilch-Warner background \cite{Dimov:2017ryz}.
The equations involving $X^5$ and $X^6$ can be combined into a single complex differential equation. Defining $Z = X^6+iX^5$, we find
\begin{equation}
\DAlambert Z + \frac{1}{2}\ \sqrt{1-6J^2}\ \big(Z-\bar Z\big)\  \partial_\sigma Z = 0
\end{equation}
This corresponds to a non-linear complex harmonic oscillator for which the exact analytic solutions can't be obtained. However, for small value of 
$\sqrt{1-6J^2}$ we can use perturbation theory to obtain the frequencies of the oscillating modes.

\section{The Non-Abelian T-dual of the Klebanov-Witten Background}

In this section, we will discuss the non-Abelian T-duality of $AdS_5\times T^{1,1}$ background. This background arises upon placing
$D3$-branes at the tip of a conifold. The field theory dual has been constructed by Klebanov and Witten \cite{Klebanov:1998hh},  \cite{Klebanov:1999tb}.
The non-Abelian T-duality
on a subgroup of the symmetry group of the internal manifold $T^{1,1}$ has been carried out in \cite{Itsios:2013wd,Barranco:2013fza,Kooner:2014cqa}.\footnote{See also \cite{Zacarias:2014wta}
for a detailed discussion on some classical sting solutions of the non-Abelian T-dual background.} An extensive
study of this T-dual background was carried out  \cite{Araujo:2015npa}. Unlike the $AdS_5\times S^5$ case, here the non-Abelian T-duality
preserves all the supersymmetries of the original background \cite{Hassan:1999bv,Itsios:2012dc,Jeong:2013jfc,Kelekci:2014ima}.\footnote{We are grateful to N. T. Macpherson
for explaining us the susy conditions along with providing us appropriate references.}
In the following we will briefly review the dual background.
Subsequently, we will discuss the Penrose limits along various null geodesics of the resulting geometry.

%\subsection{The non-Abelian T-dual solution}

The T-dual solution that we consider here  has been studied in detail in \cite{Itsios:2013wd,Barranco:2013fza,Kooner:2014cqa,Araujo:2015npa,Itsios:2012zv,Macpherson:2014eza}.
The background geometry is specified by the metric
\begin{equation}
d\hat{s}^2= L^2 ds^2_{AdS_{5}} + L^2 d\hat{s}^2_{T^{1,1}} \ ,
\end{equation}
with
\begin{equation}
d\hat{s}^2_{T^{1,1}}=\lambda_{1}^2 d\Omega_{2}^2\big(\theta_{1},\phi_{1}\big) + \frac{\lambda_{2}^2 \lambda^2}{\Delta} x_{1}^2 \sigma_{\hat{3}}^{2} + \frac{1}{\Delta} \Big[\big(x_{1}^2 + \lambda^2 \lambda_{2}^2\big) dx_{1}^2 + \big(x_{2}^2 + \lambda_{2}^4\big) dx_{2}^2 +2x_{1}x_{2}dx_{1}dx_{2} \Big] \ ,
\end{equation}
and
\begin{equation}
\Delta=\lambda_{2}^2 x_{1}^2 + \lambda^2(x_{2}^2+\lambda_{2}^{4}) \ , \  \sigma_{\hat{3}}= d\psi + \cos \theta_{1} d\phi_{1} \ .
\end{equation}
Here we have done appropriate rescaling of the coordinates $x_1$ and $x_2$ in order to get an overall factor of $L^2$ in the metric.
The NS-NS two-form of the dual background is given by the expression
\begin{equation}
\hat{B}_{2}= - \frac{\lambda^2 L^2}{\Delta}\Big[x_{1}x_{2}dx_{1} + \Big(x_{2}^2+ \lambda_{2}^{4}\Big)dx_{2}\Big] \wedge \sigma_{\hat{3}} \ ,
\end{equation}
along with the dilaton
\begin{equation}
e^{-2\hat{\Phi}}= \frac{8L^6}{g_{s}^2}\Delta \ .
\end{equation}
The corresponding NS-NS three form flux is given by
\begin{eqnarray}
\hat{H}_{3}&=&\frac{\lambda^2 L^2}{\Delta^2} \Big[\lambda_{2}^2 x_{1}^3 + \lambda^2 x_{1}\Big(x_{2}^2+\lambda_{2}^{4}\Big)-2\lambda^2 x_{1}x_{2}^2 + 2\lambda_{2}^2 x_{1}\Big(x_{2}^2+\lambda_{2}^{4}\Big)\Big] dx_{1}\wedge dx_{2} \wedge \sigma_{\hat{3}}  \cr
&-&\frac{\lambda^2 L^2}{\Delta} \Big[x_{1}x_{2} dx_{1} + \Big(x_{2}^2+\lambda_{2}^{4}\Big) dx_{2}\Big]\sin\theta_{1} d\theta_{1} \wedge d\phi_1 \ .
\end{eqnarray}

The RR sector of the background is described by the field strengths
\begin{equation}
\hat{F}_{2}=\frac{8\sqrt{2}}{g_{s}} \lambda \lambda_{1}^4 L^4 \sin\theta_{1} d\phi_{1} \wedge d\theta_{1} \ ,
\end{equation}
and
\begin{equation}
\hat{F}_{4}=-\frac{8\sqrt{2}}{g_{s}} L^6 \lambda \lambda_{1}^4 \frac{x_{1}}{\Delta} \sin\theta_{1} d\phi_{1} \wedge d\theta_{1} \wedge \sigma_{\hat{3}} \wedge \Big(\lambda_{1}^2 x_{1} dx_{2}- \lambda^2 x_{2} dx_{1}\Big) \ .
\end{equation}
It has been shown that \cite{Itsios:2013wd,Itsios:2012zv} this background solves the type IIA supergravity equations preserving
 $\mathcal{N}=1$ supersymmetry.

We will now focus on the Penrose limits around various null geodesics of the above dual geometry. We consider motion along the
isometry directions $\phi_1$ and $\psi$. Let us first focus on $\phi_1$-isometry. The relevant metric component is
\begin{equation}
g_{\phi_{1}\phi_{1}}=L^2\ \Big[\lambda_{1}^2 \sin^{2}\theta_{1} + \frac{\lambda_{2}^{2}\lambda^2}{\Delta} x_{1}^2 \cos^{2}\theta_{1}\Big] \ .
\end{equation}
This component has non-trivial dependence on $x_1,x_2$ and $\theta_1$. The geodesic condition $\partial_\mu g_{\phi_1\phi_1} = 0$
gives $x_1=0,\theta_1=\pi/2$ for $\mu=x_1$, and $x_1=0=x_2,\theta_1=\pi/2$ for $\mu=x_2$. For the choice $\mu = \theta_1$ this gives rise
to the values $\theta_1 = (0,\pi/2,\pi)$. Clearly the only non-singular choice for a geodesic is $x_{1}=0, \ x_{2}=0$ and $\theta_{1}=\pi/2$.
We will make the following large $L$ expansion around this geodesic:
\begin{equation}
r=\frac{\bar{r}}{L}, \ x_{1}=\frac{y_{1}}{L}, \ x_{2}=\frac{y_{2}}{L}, \ \theta_{1}=\frac{\pi}{2} + \frac{z}{L}, \ t=ax^{+}, \ \phi_{1}=bx^{+} + \frac{x^{-}}{L^2} \ ,
\end{equation}
while keeping the $\psi$-coordinate unchanged. The parameters $a$ and $b$ are chosen to be $1/\lambda_1$ and $1/\lambda_1^2$ respectively. Further, we
redefine the coordinates as  $x^{+}=u,\ x^{-}=v$ and rescale $z\rightarrow\sqrt{6}z, \ y_{1}\rightarrow y_1/\sqrt{6},\ y_{2}\rightarrow y_{2}/3$.
The leading order terms of the metric in the limit $L\rightarrow \infty$ gives
\begin{equation}\label{natdppw}
ds^2= 2dudv + d\bar{r}^2 + \bar{r}^2 d\Omega_{3}^{2} + dz^2 + dy_{1}^2 + y_{1}^2 d\psi^2 + dy_{2}^2 - 6 \big(\bar{r}^2+ 6 z^2\big)du^2.
\end{equation}
This is indeed a pp-wave solution in the standard Brinkmann form. Interestingly, the pp-wave metric in the above is identical to the metric \eqref{bgppwave}, 
we have obtained from the Abelian T-dual background.

We will now focus on other background fields. In order to keep the dilaton finite, we redefine the string coupling as
\begin{equation}
g_{s}=L^{3} \tilde{g_{s}} \ .
\end{equation}
With this redefinition, the dilaton takes the form
\begin{equation}
e^{-2\hat{\Phi}}=\frac{8}{\tilde{g}_{s}^2} \lambda^2 \lambda_{2}^4 \ ,
\end{equation}
In this limit, the NS-NS two-form field on the other hand becomes
\begin{equation}
\hat{B}_{2}= 2\sqrt{6} z \ dy_{2} \wedge du \ ,
\end{equation}
with the corresponding three-form flux
\begin{equation}
\hat{H}_{3}= 2\sqrt{6} \ du\wedge dz \wedge dy_{2} \ .
\end{equation}
The RR fields at Penrose limit are given as
\begin{equation}
\hat{F}_{2}=\frac{8}{3\sqrt{3}\tilde{g}_{s}} \ du \wedge dz, \ \hat{F}_{4}=0 \ .
\end{equation}

The motion on along the $\psi$-isometry however does not give pp-wave geometry as we will see in the following. The relevant component
of the metric is
\begin{equation}
g_{\psi\psi}= L^2 \ \frac{\lambda_{2}^2\lambda^2}{\Delta} x_{1}^2\ .
\end{equation}
From the above we obtain the geodesic $x_2=0, \theta_1 = 0$. Consider the following expansion
\begin{equation}
r=\frac{\bar{r}}{L}, \ x_{2}=\frac{y_{2}}{L}, \ \theta_{1}=\frac{z}{L},\ t=ax^{+}, \ \psi= bx^{+}+ \frac{x^-}{L^2}\ ,
\end{equation}
while keeping $x_{1}$ and $\phi_{1}$ coordinates unchanged. The leading terms of the dual metric in $L\rightarrow\infty$ becomes
\begin{eqnarray}
ds^2 &=& -\bar{r}^2 a^2 (dx^{+})^2+ d\bar{r}^2 + \bar{r}^2 d\Omega_{3}^2 + \lambda_{1}^2 dz^2 + \lambda _{1}^2 z^2 d\phi_{1}^2\nonumber \\
 &+& \frac{\lambda_{2}^2\lambda^2}{\sum} \Big[2b x_{1}^2 \ dx^+ dx^- + 2x_{1}^2 dx^- d\phi_1- bx_{1}^2z^2dx^+ d\phi_1 -z^2 x_{1}^2d\phi_{1}^2\nonumber\\
 &-&\frac{\lambda^2 y_{2}^2 x_{1}^2}{\sum} \big(bdx^+ + d\phi_1\big)^2\Big] - \frac{1}{\sum} \Big[\frac{1}{\sum}\lambda^2 y_{2}^2 \big(x_{1}^2 + \lambda_{2}^2\lambda^2\big)dx_{1}^2 - \lambda_{2}^4 dy_{2}^2 - 2x_1 y_2 dx_1 dy_2 \Big]\nonumber\\
 &- &  L^2 a^2 (dx^{+})^2 +\frac{L^2}{\sum} \Big[\lambda_{2}^2\lambda^2x_{1}^2 \left\{  \big(bdx^+ + d\phi_1\big)^2 +
2b dx^+ d\phi_1 \right\} + \big(x_{1}^2 + \lambda_{2}^2\lambda^2\big)dx_{1}^2 \Big] \
 \end{eqnarray}
where,
\begin{equation}
\sum=\lambda_{2}^2 x^2_{1} + \lambda^2 \lambda^4_{2}\ .
\end{equation}
In this case too the geodesic is not null for any choice of the parameters $a$ and $b$. This is reflected by the appearance of the divergent term in the metric. 
Hence the motion along $\psi$-isometry does not give pp-wave geometry.

\section{Closed String Quantization on the PP Wave}

We will now study the quantization of a closed string propagating in the pp-wave background \eqref{natdppw}, derived in the last section. The worldsheet action is given by
\begin{equation}
S=-\frac{1}{4\pi\alpha^{\prime}}\int d\tau d\sigma \Big[\sqrt{g} g^{\alpha\beta}G_{\mu\nu} \partial_{\alpha}X^{\mu}\partial_{\beta}X^{\nu} + \epsilon^{\alpha\beta}B_{\mu\nu} \partial_{\alpha}X^{\mu}\partial_{\beta}X^{\nu} + \alpha^{\prime} \sqrt{g}\ \mathcal{R}^{(2)} \Phi\Big] \ ,
\end{equation}
As before, we will use the notation $\epsilon^{\tau\sigma}=-\epsilon^{\sigma\tau}=1$ and gauge fix the metric as
$\sqrt{g} g^{\alpha\beta}=\eta^{\alpha\beta}$ with the convention $ -\eta_{\tau\tau}=\eta_{\sigma\sigma}=1$.
We assign string coordinates as
\begin{equation}
U=u,\ V=v, \ \Big(X^1,X^2,X^3,X^4\Big)\in \bar{r},\Omega_{3},\ \Big(X^5,X^6\Big)\in y_{1},\psi, \ \Big(X^7,X^8\Big)\in z,y_{2} \ .
\end{equation}
Further, we fix the residual diffeomorphism invariance considering  the light cone gauge $U=\tau$  with $p^+=1$.
The worldsheet action for the pp-wave background \eqref{natdppw}  becomes
\begin{equation}
S=-\frac{1}{4\pi\alpha^{\prime}}\int d\tau d\sigma\ \Big[\partial X^{i}.\ \partial X^{i}
+6\ \Big(\sum_{i=1}^{4} (X^i)^2 + 6 (X^7)^2\Big) - \sqrt{6}X^7 \partial_{\sigma}X^8 + \sqrt{6}X^8 \partial_{\sigma}X^7\Big] \ .
\end{equation}

The equations of motion for the scalar fields in the above action are given by
\begin{equation}
\DAlambert X^{i}-6X^{i}=0,\\ i=1,2,3,4,
\end{equation}

\begin{equation}
\DAlambert X^{i}=0, \\ i=5,6,
\end{equation}

\begin{equation}
\DAlambert X^{7}-36 X^7 + \frac{1}{2} \sqrt{6}\ \partial_{\sigma} X^{8}=0,
\end{equation}

\begin{equation}
\DAlambert X^{8}-\frac{1}{2} \sqrt{6}\ \partial_{\sigma} X^{7}=0.
\end{equation}
To obtain the oscillator frequencies we consider an ansatz of the form $X^{i}\sim e^{-i\omega t + in \sigma}$. We find

\begin{equation}
\omega^{2}_{n,i}=n^2 + 6, \\ i=1,2,3,4,
\end{equation}

\begin{equation}
\omega^{2}_{n,i}=n^2, \\ i=5,6,
\end{equation}

\begin{equation}
\omega^{2}_{n,i}=n^2 + \frac{1}{2} \Big[36\pm \sqrt{(36)^2+6n^2} \ \Big], \\ i=7,8.
\end{equation}

\section{Field Theory Duals}

In the previous discussion we have seen that taking the Penrose limit gives rise to pp wave geometries for smooth null geodesics, both in the case of Abelian as well as 
non-Abelian T-dual backgrounds from $AdS_5\times T^{1,1}$. Here we will discuss the underlying field theory duals. We will first consider the Abelian T-duals. 
The field theory duals for these backgrounds has been constructed in  \cite{Uranga:1998vf} and  \cite{Dasgupta:1998su}. They correspond to a system of intersecting
$D4-NS5-NS5'$ branes where the $NS5$ branes are rotated appropriately and the $D4$ branes are stretched in between. Here we will study the field theory dual of the 
corresponding pp-wave geometries.

In the limit of large $F_5$ flux, the string coupling becomes weak and hence the pp-wave background can be treated semi-classically. To show this, note that the type 
$IIA$ and type $IIB$ string couplings $g_s^A$ and $g_s^B$ are related among each other as $g_{s}^B=g_{s}^A L$. In order to get a finite dilaton in the Penrose limit 
we have rescaled the string coupling of the T-dual geometry as $\tilde g_s = g_s^A/L$. Hence, the $IIB$ string coupling is related to $\tilde g_s^A$ as $g_s^B = L^2 \tilde g_s$.
For the Klebanov-Witten background the size $L$ of $AdS_5$ space  is quantized in terms of the $F_5$ flux $N_3$ as \cite{Itsios:2017cew}:
\begin{equation}\label{ksq}
L^4 = \frac{27}{4} \pi {g_{s}^B}^2 N_3 \ .
\end{equation}
Thus we find 
\begin{equation}
\tilde{g}_s \sim \frac{1}{\sqrt{N_3}} \ .
\end{equation}
This shows that, in the limit of large $N_3$ the string coupling $\tilde g_s$ becomes negligible. Thus, it seems plausible to use semi-classical analysis to compute
the spectrum for our purpose.

The construction of the field theory dual is as follows  \cite{Uranga:1998vf,Dasgupta:1998su}. It describes the dynamics of massless strings arising from $N$-D4 branes 
stretched across two orthogonal NS5 branes located on a circle. The spectrum consists of two chiral multiplets $A_1, A_2$ in the $(N,\overline N)$ 
representation and two more chiral multiplets $B_1,B_2$ in the $(\overline N,N)$ representation of the $SU(N)\times SU(N)$ gauge group, with superpotential 
\begin{eqnarray}
W= \frac{1}{2} \epsilon^{ij} \epsilon^{kl} \ Tr \big[ A_i B_k A_j B_l \big] \ , \ \ i,j,\ldots = 1,2. 
\end{eqnarray}
There is an underlying $SU(2)_A\times SU(2)_B\times U(1)_R$ global symmetry preserved by the theory. The fields $(A_1,A_2)$ form a doublet under the $SU(2)_A$ subgroup 
of the global symmetry and similarly $(B_1,B_2)$ form a doublet under $SU(2)_B$. The R-symmetry $U(1)_R$ originates due to a shift along the circle coordinate, and 
all the fields $A_1,A_2,B_1,B_2$ transform by the same phase under this symmetry. Let us denote $J_1$ and $J_2$ to be the Cartan generators of $SU(2)_A$ and $SU(2)_B$
respectively and let $J_3$ be the generator of $U(1)_R$. 

This field theory system is dual to the T-dual geometry specified by the metric \eqref{abelianTdual}: 
\begin{eqnarray*}
d\hat{s}^2 = L^2 ds^2_{AdS_{5}} + L^2\ \Big[\lambda_{1}^2 d\Omega_{2}^2\big(\theta_{1},\phi_{1}\big)
+ \lambda_{2}^2 d\Omega_{2}^2\big(\theta_{2},\phi_{2}\big) + \frac{1}{\lambda^2} d\psi^2\Big] \ .
\end{eqnarray*}
The $SU(2)_A$ and $SU(2)_B$ are identified with the symmetries of the two spheres parametrized by $(\theta_1,\phi_1)$ and $(\theta_2,\phi_2)$ and the R-symmetry $U(1)_R$
is identified with the shift along the $\psi$-direction. The generators $J_1$ and $J_2$ correspond to the shift in the azimuthal coordinates $\phi_1$ and $\phi_2$ respectively and 
$J_3$ corresponds to shift in $\psi$. The BMN sector for the field theory dual has been constructed \cite{PandoZayas:2002dso}. The state operator correspondence is naturally 
described in terms of the conifold coordinates $Z_1 = A_1B_1, Z_2 = A_2B_2, Z_3 = A_1B_2, Z_4 = A_2B_1$. Setting the light cone Hamiltonian $H = \Delta-(J_1+J_2+J_3)$, it 
can be shown that the operator $Z_1$ has $H=0$ and corresponds to the ground state. The first excited state, with $H=1$ is described in terms of the operators 
$Z_3, Z_4$ and  the covariant derivatives $D_iZ_1$.

This is in contrast to what we observe in closed string quantization of the pp-wave geometry \eqref{ppwave1}, corresponding to the Abelian T-dual background.
From  \eqref{eigenfreq} we find the frequencies corresponding to $n=0$ modes as\footnote{We need not worry about the modes corresponding to the non-linear oscillators here. For small value of $\sqrt{1-6J^2}$ it can be shown using perturbation theory 
that the lowest mode will correspond to $n=1$ and will have a higher frequency than the above modes. }
\begin{eqnarray}
&& \omega_{0,i}= 1, \ i=1,2,3,4, \nonumber\\
&& \omega_{0,7+,8+}= 6J\ , \nonumber\\
&& \omega_{0,7-,8-}= 0 \ .
\end{eqnarray}
This mismatch, however, is not surprising. Large effective interaction cause the energies of the states to change. A similar phenomenon has been observed for the pp wave
background corresponding to the Abelian T-dual of $AdS_5\times S^5$ geometry \cite{Itsios:2017nou}.

For the non-Abelian T-dual background of $AdS_5 \times T^{1,1}$, the field theory dual have been first proposed in \cite{Itsios:2013wd} and subsequently, with suitable 
modification, analysed extensively in \cite{Itsios:2017cew}. The dual theory is conjectured to  arise from an intersecting $D4-NS5-NS5'$ brane configuration.  Due to Myers 
effect, the $D4$ branes are blown into a stack of $D6$ branes on a sphere in the presence of $B_2$ field. The $NS5$ and $NS5'$ branes are located at various points on the 
radial direction and are transverse to two different $S^2$s. One $NS5'$ brane is placed between two consecutive $NS5$ branes. Due to large gauge transformation, the 
$D4$ brane charge changes by a fixed amount each time a $NS5$ brane is crossed. The dual theory consists of a two tailed linear quiver with gauge groups of increasing 
rank at each node and matter fields in the bifundamental of each pair of nodes with a suitably added flavour group at the middle. The holographic dual corresponding to 
the pp-wave geometry resulting from the non-Abelian T-dual background will correspond to a class of operators in this quiver theory. Note that  the construction of the field 
theory dual in \cite{Itsios:2017cew} was mainly based on the brane charges arising from the supergravity background and the scaling of the central charge. The central 
charge corresponding to the quiver theory was computed using $a$-maximisation and was shown to agree with the holographic entanglement entropy computed from the 
background associated with the supergravity dual \cite{Itsios:2017cew}. Naively one might expect that a similar analysis can also be carried out for the corresponding 
pp-wave background. However, care must be taken in the present case because the pp-wave geometry is obtained in zooming a particular region and hence it is globally 
{\it not complete} \cite{Itsios:2017cew}. In this case, the holographic entanglement entropy can be computed, as in \cite{Macpherson:2014eza} by imposing a hard cutoff 
on the non-compact directions. However the field theory interpretation of this quantity is not clear. It might correspond to the entanglement entropy in some excited state 
of the dual field theory.

\section{Conclusion}

In this paper we have studied the Abelian as well as non-Abelian T-dual geometries arising from the Klebanov-Witten background. Though the Abelian T-duality is an exact 
symmetry of the string theory, the dual description is some times more convenient to study the Penrose limits. The non-Abelian T-duality provides new supergravity solutions. We 
considered various null geodesics of the resulting dual theories and obtained the Penrose limits. Some of these geodesics are singular while the remaining admit pp-wave geometries.  We quantized closed strings propagating in these pp-wave backgrounds. We have briefly analysed the corresponding field theory duals. For the non-Abelian 
case the holographic dual of the pp-wave geometry corresponds to a sector of operators of a quiver theory with gauge groups of increasing rank. Further investigation is 
required to identify this BNM sector and to establish a precise mapping between holographically computed quantities and field theory observables. It would also be interesting 
to explore the possibility of obtaining pp-wave geometries for non-Abelian duals of string theory compactified on $T^{p,q}$ as well as backgrounds with $AdS_3$ factors. We 
hope to report on some of these issues in future. 

\vskip .3in
 
\noindent {\bf\large Acknowledgement}

 \vskip .2in

\noindent
The present work  is  partially supported by the DST project grant  no. EMR/2016/001997.

\vskip .2in

\noindent{\bf\large A. Bulk Modes} \\

The bulk modes play an important role in obtaining the spectrum of the dual field theory. 
In order to understand the holographic dual of the pp-wave geometry obtained from the non-Abelian T-dual of $AdS_5\times T^{1,1}$
 we will consider a non-interacting, massless scalar field in this background and obtain the corresponding bulk modes. The pp-wave metric we are interested in is
\begin{equation}
ds^2_{pp}= 2dudv + d \bar{r}^2 + \bar{r}^2 d\Omega_{3}^2 + dz^2 + dy_{1}^2 + y_{1}^2 d\psi^2 + dy_{2}^2 -6\big(\bar{r}^2 + 6z^2\big)du^2 \ .
\end{equation}
%At large $\bar{r}$, the geometry becomes $S^3 \times R$ coordinatized by $\Omega_{3}, u$. Hence, the holographic coordinate is $\bar{r}$.
In order to obtain the bulk modes we will rewrite the above metric in a convenient form. We assign coordinates $X^i, i=1,\ldots,4$ to parametrize the
$\mathbb{R}^4$ part $\{\bar r,\Omega_3\}$, relabel the $\mathbb{R}^2$ factor parametrized by $\{y_1,\psi\}$ as $\{X^5,X^6\}$ and the
$\mathbb{R}^2$ factor in $\{z,y_2\}$ as $\{X^7,X^8\}$ while leaving the light cone coordinates $\{u,v\}$ intact. The metric now takes the form
\begin{equation}
ds^2_{pp}= 2dudv- 6\ \Big[\sum_{i=1}^4 (X^i)^2 + 6 (X^7)^2\Big]du^2 + \sum_{i=1}^8 \big(dX^i\big)^2 \ .
\end{equation}
The NS-NS and RR fields in this coordinate system are given by
\begin{equation}
\hat{B}_{2}= 2\sqrt{6} X^7 \ dX^8 \wedge du \ ,\ e^{-2\hat{\Phi}}=\frac{8}{\tilde{g}_{s}^2} \lambda^2 \lambda_{2}^4 \ ,
\end{equation}
and
\begin{equation}
\hat{F}_{2}=\frac{8}{3\sqrt{3}\tilde{g}_{s}} \ du \wedge dX^7 \ , \ \hat{F}_{4}=0 \ .
\end{equation}

This geometry preserves $SO(4)\times SO(2)\times U(1)$ symmetry. Rotations in $X^1,X^2,X^3,X^4$ gives rise to the $SO(4)$ factor and
rotations in $X^5,X^6$ gives rise to the $SO(2)$. There is an additional translational symmetry giving rise to $U(1)$ symmetry. Note that,
in contrast the background $AdS_5\times T^{1,1}$ has $SO(2,4)\times SU(2)\times SU(2)\times U(1)$ symmetry whereas the non-Abelian
T-dual geometry possesses $SO(2,4)\times SU(2)\times U(1)$. Taking Penrose limit this symmetry reduces to $SO(4)\times SO(2)\times U(1)$.

We consider a massless scalar field $\Phi$ in this background for which the equation of motion is given by
\begin{equation}
\Box \Phi=0 \ .
\end{equation}
with the Laplacian
\begin{equation}
\Box= 2 \partial_{u} \partial_{v} + 6\ \Big[\sum_{i=1}^4 (X^i)^2 + 6 (X^7)^2\Big] \partial_{v}^2 + \sum _{i=1}^8 \partial_{X^i}^2 \ .
\end{equation}
To solve this wave equation, we use the method of separation of variables. Set
$$ \Phi(u,v,X^i) = f(u,v) g(X^i) \ . $$
The wave equation gives
\begin{eqnarray}
\frac{2\partial_u\partial_vf}{f(u,v)} + 6 \Big[\sum_{i=1}^4 (X^i)^2 + 6 (X^7)^2\Big] \frac{\partial_v^2f}{f(u,v)} + \frac{\Box_8 g(X^i)}{g(X^i)} = 0 \ .
\end{eqnarray}
Setting the ansatz $f(u,v) \sim e^{i(p_v v - p_uu)}$, we obtain
\begin{equation}
\Box_8g(X^i) - 6 p_v^2\Big[\sum_{i=1}^4 (X^i)^2 + 6 (X^7)^2\Big] g(X^i) + 2 p_up_v g(X^i) = 0 \ .
\end{equation}
This equation now has a familiar Harmonic oscillator form whose solutions are given in terms of well known Hermite polynomials. We find
\begin{equation}
\Phi (u,v, X^i)= e^{i\big(p_v v - p_uu + c_5X^5 + c_6X^6+c_8X^8\big)} \ e^{-\frac{\beta X_{7}^2}{2}} H_{n_7} \big(\sqrt{\beta} X^7\big)\ \prod_{j=1}^4 e^{-\frac{\alpha X_{j}^2}{2}} H_{n_{j}} \big(\sqrt{\alpha} X^j\big) \ ,
\end{equation}
Here $p_{u},p_{v}$ are the conserved canonical momenta along $u$ and $v$ direction respectively. For convenience we have used the notation
$\alpha^2=6p_{v}^2, \beta^2=36p_{v}^2, \sum_{i=1}^8c_i^2 = 2 p_up_v,$ and $n_i = (c_i^2\alpha - 1)/2$ in the above expression.
% $H_{n}$'s are Hermite polynomials which are normalizable.\\

%\newpage

\end{document}